\documentclass[12pt]{article}
\pdfoutput=1
\usepackage{subfigure}
\usepackage{ulem}
\usepackage{amssymb,amsmath}
\usepackage{graphicx}
\usepackage{color}
\usepackage[colorlinks=true
,urlcolor=blue
,citecolor=blue
,linkcolor=blue
,linktocpage=true
,pdfproducer=medialab
]{hyperref}
\usepackage[a4paper,width=15.3cm]{geometry}

\begin{document}

\begin{center}

 {\Large\bf   Top Quark and Higgs Boson Masses  in  Supersymmetric  Models
 } \vspace{1cm}

{\bf  Ilia Gogoladze$^{a,}$\footnote{E-mail:
ilia@bartol.udel.edu\\ \hspace*{0.5cm} On  leave of absence from:
Andronikashvili Institute of Physics, 0177 Tbilisi, Georgia.},  Rizwan Khalid $^{b,}$\footnote{E-mail: rizwan.hep@gmail.com, rizwan@sns.nust.edu.pk},
Shabbar Raza$^{c,}$\footnote{Email: shabbar@udel.edu}
  and   Qaisar Shafi$^{a,}$\footnote{E-mail:
shafi@bartol.udel.edu} } \vspace{.5cm}

{ \it
$^a$Bartol Research Institute, Department of Physics and Astronomy, \\
University of Delaware, Newark, DE 19716, USA  \\
$^b$Department of Physics, School of Natural Sciences, National
University of Sciences \& Technology, H-12, Islamabad, Pakistan\\
$c$
State Key Laboratory of Theoretical Physics and Kavli Institute for Theoretical Physics China (KITPC),
Institute of Theoretical Physics, Chinese Academy of Sciences, Beijing 100190, P. R. China
} \vspace{.5cm}

\vspace{1.5cm}
 {\bf Abstract}
\end{center}

We study the implications for  bounds on the top quark pole mass $m_t$ in models with low scale supersymmetry following the discovery of the
Standard Model-like Higgs boson. In the minimal supersymmetric standard model, we find that  $m_t\ge 164$ GeV,
if the light CP even Higgs boson mass $m_h = 125\pm2$ GeV. We also explore the top quark and Higgs boson masses in two classes of
supersymmetric SO(10) models with t-b-$\tau$ Yukawa coupling unification at $M_{\rm GUT}$. In particular, assuming SO(10) compatible non-universal
gaugino masses, setting $m_h = 125$ GeV and requiring 5\% or better Yukawa unification, we obtain the result $172~{\rm GeV}\leq m_t \leq 175~{\rm GeV}$.
Conversely, demanding  $5\%$ or better $t$-$b$-$\tau$ Yukawa unification and setting
 $m_t=173.2$ GeV, the Higgs boson mass is predicted to lie in the range $122~{\rm GeV}\leq  m_h \leq 126~{\rm GeV}$.

\newpage

\renewcommand{\thefootnote}{\arabic{footnote}}
\setcounter{footnote}{0}



\renewcommand{\thefootnote}{\arabic{footnote}}
\setcounter{footnote}{0}



\section{Introduction}

With  the discovery by the ATLAS \cite{:2012gk} and CMS \cite{:2012gu} collaborations
 { of} a Standard Model (SM)-like Higgs boson with mass $m_h \simeq 125-126$ GeV,  the particle  content
of the SM is fully  verified,   and the top quark with
mass $m_t = 173.18 \pm 0.94\, {\rm GeV}$  \cite{Aaltonen:2012ra} remains the heaviest particle.
The flavor structure of the SM, and, in particular, the relatively  heavy top quark mass compared with other fermion masses
is an open question in the SM. Likewise, as far as the SM is concerned, there is no deep reason for the Higgs
boson to have mass $m_h \simeq 125-126$ GeV.

 At the same time, one of the successes of the SM  is a relatively  accurate
prediction of the top quark pole  mass $m_t$ from fits to the electroweak data. The top quark `prediction' deduced  from
such  fits does not change much as a result   of the  Higgs boson discovery   due to a relatively mild
dependence on the latter~\cite{Baak:2012kk,Eberhardt:2012gv}.
This mild dependence of the electroweak fits on the Higgs boson mass can be understood from the  one loop radiative corrections  to the $W$-boson mass.
While these corrections depend quadratically on the top quark mass, there is only a mild  logarithmic
dependence on the SM Higgs boson mass.
Employing all experimental
data (including $m_h=125.7$ GeV), fits to the electroweak data at the $95\%$  confidence level yield  $168\, {\rm GeV}\lesssim m_t \lesssim 178\,{\rm GeV}$, and 
$80.3\,{\rm GeV} \lesssim M_W \lesssim 80.4 \,{\rm GeV}$~\cite{Baak:2012kk}  {for the $W$-boson mass}, in good
agreement  with
the direct measurement of these quantities.
Note that while it is possible to constrain the top quark in the SM
from rare decays, such constraints are not compatible with the bounds obtained from the determination of
the $W$ boson mass~\cite{Baak:2012kk,Eberhardt:2012gv}.

Supersymmetry remains a compelling extensions of the SM and   the top quark plays a key role in several features of the
minimal supersymmetric standard model (MSSM). For instance, the top quark Yukawa coupling is the
dominant contributor to the mechanism of radiative electroweak symmetry breaking (REWSB). We discuss  in this paper
how the REWSB condition  restricts the top quark mass  as a function of
$\tan\beta$, the ratio of the vacuum expectation values of the two MSSM Higgs doublets. The
top quark  also plays a crucial role  in the calculation of radiative corrections to  the lightest CP even Higgs boson
mass in the the MSSM. Among other things these corrections, to leading order, are proportional to the fourth power of the top quark mass.
In addition, there is a  logarithmic dependence  on the geometric mean $M_S$ of the stop quark masses, as well as
 a contribution proportional to $(A_t/M_S)^4$, where $A_t$ is {a} tri-linear soft supersymmetry breaking (SSB) term \cite{at}.
We  show in this paper that a 125 GeV SM-like Higgs boson
provides  strong constraints  on the allowed mass range for the top quark mass.  This mass
interval turns out to be  compatible  with the range obtained from fits to  the electroweak data.

{In a supersymmetric SO(10) grand
unified theory (GUT) with
 $t$-$b$-$\tau$ Yukawa unification \cite{big-422}, it was shown in \cite{Gogoladze:2011aa} that with suitable
non-universal
SSB gaugino masses at $M_{\rm GUT}$   the lightest CP even
Higgs boson mass is predicted {to} lie close to   125 GeV. Such gaugino non-universality  at $M_{\rm GUT}$ can arise, for instance,
from a non-singlet $F$-component of the field that
breaks supersymmetry. }
Motivated by this result we  seek to provide an answer in this paper to  the following question: Can models compatible with $t$-$b$-$\tau$
Yukawa unification  also yield  a stringent constraint for the top quark mass in good agreement with the observations?

We consider two classes of SO(10) models with a minimal
set of SSB parameters  at  $M_{\rm GUT}$ in which  $t$-$b$-$\tau$ Yukawa unification is realized.
The most well studied one  has universal SSB gaugino
mass terms but non-universal Higgs SSB terms, $m^2_{H_{u}}\neq m^2_{H_{d}}$ at $M_{\rm GUT}$.
Here $m^2_{H_{u},H_d}$  stand for the up/down type Higgs SSB masses$^2$.
The second  class of models, mentioned earlier,   {assumes} universal Higgs SSB mass$^2$ terms,  but the  gaugino
SSB masses at $M_{\rm GUT}$ are non-universal.
Allowing  the top quark mass  to vary in the interval $0< m_t<220$ GeV,
we scan the characteristic SSB parameter space for both  classes  of models
and show that  the $t$-$b$-$\tau$ Yukawa unification condition yields  \sout{to} a relatively narrow
interval for the masses of the top quark and the light CP even  Higgs boson. An upper bound for the top quark mass is obtained by
 imposing perturbativity on  the top Yukawa coupling up to $M_{\rm GUT}$.
It is possible \cite{Gogoladze:2009ug,Badziak:2012mm} to relax the universality of SSB Higgs  mass$^2$ terms along with the
non-universality of SSB gaugino  mass terms,
but we restrict ourselves here to the above two classes.

The outline for the rest of the paper is as follows. In section 2 we summarize the scanning procedure and
the experimental constraints applied in our analysis. In Section 3 we present the results for the SO(10)
model in which  $t$-$b$-$\tau$ Yukawa unification is  achieved  via  non-universal SSB gaugino masses
and universal up and down Higgs SSB mass$^2$ terms ($m^2_{H_{u}}=m^2_{H_{d}}=m^2_{10}$). Using the fact that
the light CP even Higgs boson mass is in the interval  $123~{\rm GeV} < m_h < 127~{\rm GeV}$, we show that
 the top quark mass lies  in the range  {$164\, {\rm GeV}< m_t< 205~{\rm GeV}$}. Imposing $5\%$ or better
  $t$-$b$-$\tau$ Yukawa unification,    the allowed top quark mass range shrinks   to $168~{\rm GeV}<m_t<180~{\rm GeV}$. In
Section 4 we present results for  the SO(10)  model with  universal SSB gaugino mass terms but which also requires that
$m_{H_{u}}^2< m_{H_{d}}^2$ \cite{Blazek:2002ta}. In
this case, we find the top quark mass range   {$165~{\rm GeV}< m_t< 200~{\rm GeV}$}.  Imposing $5\%$ or better $t$-$b$-$\tau$
Yukawa unification condition, the interval for the top quark mass is reduced  to
\textbf {$170~{\rm GeV}<m_t<178~{\rm GeV}$}. We also consider $b$-$\tau$ Yukawa unification in this case and, as expected,
the constraint on the top quark mass is  relaxed to  $168~{\rm GeV}< m_t<200~{\rm GeV}$. Our conclusions are presented in Section 5.

\section{Phenomenological constraints and scanning procedure \label{pheno}}
\label{sec:scan}
We employ the ISAJET~7.84 package~\cite{ISAJET}
to perform random scans over the parameter space.
In this package, the weak scale values of gauge and third
generation Yukawa couplings are evolved to
$M_{\rm GUT}$ via the MSSM renormalization group equations (RGEs)
in the $\overline{DR}$ regularization scheme.
We do not strictly enforce the unification condition
$g_3=g_1=g_2$ at $M_{\rm GUT}$, since a few percent deviation
from unification can be assigned to unknown GUT-scale threshold
corrections~\cite{Hisano:1992jj}.
With the boundary conditions given at $M_{\rm GUT}$,
all the SSB parameters, along with the gauge and third family Yukawa couplings,
are evolved back to the weak scale $M_{\rm Z}$.

In evaluating  the Yukawa couplings the SUSY threshold
corrections~\cite{Pierce:1996zz} are taken into account
at a common scale  $M_S= \sqrt{m_{\tilde t_L}m_{\tilde t_R}}$.
The entire parameter set is iteratively run between
$M_{\rm Z}$ and $M_{\rm GUT}$ using the full 2-loop RGEs
until a stable solution is obtained.
To better account for the leading-log corrections, one-loop step-beta
functions are adopted for the gauge and Yukawa couplings, and
the SSB scalar mass parameters $m_i$ are extracted from RGEs at appropriate scales
$m_i=m_i(m_i)$.The RGE-improved 1-loop effective potential is minimized
at an optimized scale  $M_S$, which effectively
accounts for the leading 2-loop corrections. Full 1-loop radiative
corrections are incorporated for all sparticle masses.

In scanning the parameter space, we employ the Metropolis-Hastings
algorithm as described in \cite{Belanger:2009ti}.
The data points collected all satisfy the requirement of radiative electroweak symmetry breaking
(REWSB)~\cite{Ibanez:1982fr},
with the neutralino in each case being the LSP.
After collecting the data, we impose the mass bounds on
all the particles \cite{Nakamura:2010zzi} and
use the IsaTools package~\cite{Baer:2002fv} and SuperIso v2.3 \cite{Mahmoudi:2008tp}
to implement the following phenomenological constraints:
\begin{table}[h!]\centering
\begin{tabular}{rlc}
$ 0.8 \times 10^{-9} \leq BR(B_s \rightarrow \mu^+ \mu^-) $&$ \leq\, 6.2 \times 10^{-9} \;
 (2\sigma)$        &   \cite{:2007kv}      \\
$2.99 \times 10^{-4} \leq BR(b \rightarrow s \gamma) $&$ \leq\, 3.87 \times 10^{-4} \;
 (2\sigma)$ &   \cite{Barberio:2008fa}  \\
$0.15 \leq \frac{BR(B_u\rightarrow
\tau \nu_{\tau})_{\rm MSSM}}{BR(B_u\rightarrow \tau \nu_{\tau})_{\rm SM}}$&$ \leq\, 2.41 \;
(3\sigma)$ &   \cite{Barberio:2008fa}  \\
 $ 0 \leq \Delta(g-2)_{\mu}/2 $ & $ \leq 55.6 \times 10^{-10} $ & \cite{Bennett:2006fi}
\end{tabular}\label{table}
\end{table}

We also implement the following  mass bounds on the sparticle masses:
\begin{table}[h!]\centering
\begin{tabular}{rlc}
 $m_{\tilde{g}} \gtrsim  1.4~{\rm TeV}~ ({\rm for}~ m_{\tilde{g}}\sim m_{\tilde{q}})$ &~\cite{Aad:2012fqa,Chatrchyan:2012jx}\\
 $m_{\tilde{g}}\gtrsim 0.9~{\rm TeV}~ ({\rm for}~ m_{\tilde{g}}\ll
m_{\tilde{q}})$ &~\cite{Aad:2012fqa,Chatrchyan:2012jx} \\
$M_A \gtrsim 700~{\rm GeV}$~ $({\rm for}$~ $\tan\beta\simeq 48$) & ~\cite{cms-mA}
\end{tabular}\label{table2}
\end{table}

\section{SO(10) GUT with non universal gauginos masses}

One of the main motivations of supersymmetric SO(10) GUT, in addition to gauge
coupling unification, is matter unification.
The spinor representation of SO(10) unifies all matter fermions of a
given family in a single multiplet ($16_i$), which also contains the right handed neutrino ($\nu_R$).
Another virtue of SO(10) is that, in principle, the two MSSM Higgs doublets can be
accommodated in a single ten dimensional ($10_H$) representation, which then yields the following Yukawa couplings
\begin{align}
Y_{ij}\ 16_i\, 16_j\, 10_{\rm H}.
\label{10-yukawa}
\end{align}
Here $i,j=1, 2, 3$ stand for family indices and the SO(10) indices have been omitted for simplicity.
For the third generation quarks and  leptons, the interaction in
Eq.(\ref{10-yukawa}) yields  the following Yukawa coupling unification condition at
 $M_{\rm GUT}$~\cite{big-422}
\begin{align}
Y_t = Y_b = Y_{\tau} = Y_{\nu_{\tau}}. \label{f1}
\end{align}
In  gravity  mediated supersymmetry  breaking
scenario \cite{Chamseddine:1982jx}, implementing Eq. (\ref{f1}),
in particular $Y_t = Y_b = Y_{\tau} $  at $M_{\rm GUT}$, can
place significant constraints on the supersymmetric spectrum \cite{susy-thres}.
These constraints depend on the particular boundary conditions for sparticle SSB masses
chosen at  $M_{\rm GUT}$.
If  the SSB gaugino mass terms are assumed to be universal at $M_{\rm GUT}$,
 ${m^2_{H_{u}}}(M_{\rm GUT}) < {m^2_{H_{d}}} (M_{\rm GUT})$  is required in order  for
Yukawa coupling unification to be consistent with radiative electroweak symmetry breaking (REWSB).
This   splitting may arise, for example, via a $D$-term contribution to all
scalar masses, or it can be generated via ``Just-So'' splitting \cite{Blazek:2002ta}.
The results of this scenario are presented in Section 4.

Alternatively, it is possible to achieve $t$-$b$-$\tau$ Yukawa coupling unification
consistent with REWSB by assuming the gaugino SSB mass terms to be non-universal at $M_{\rm GUT}$.
In this case,  non-universality among  the Higgs SSB mass
terms is not needed.

It has been pointed out~\cite{Martin:2009ad} that non-universal MSSM gaugino masses
at $ M_{\rm GUT}$ can arise from non-singlet $F$-terms, compatible with the underlying GUT symmetry.
The SSB gaugino masses in supergravity~\cite{Chamseddine:1982jx} can arise, say, from the following
dimension five operator:
\begin{align}
 -\frac{F^{ab}}{2 M_{\rm
P}} \lambda^a \lambda^b + {\rm c.c.}
\label{eq4}
\end{align}
Here $\lambda^a$ is the two-component gaugino field, $ F^{ab} $ denotes the $F$-component of
the field which breaks SUSY, and  the indices $a,b$ run over
the adjoint representation {of the gauge group}. The resulting gaugino
mass matrix is $\langle F^{ab} \rangle/M_{\rm P}$, where the
supersymmetry breaking  parameter $\langle F^{ab} \rangle$
transforms as a singlet under the MSSM gauge group $SU(3)_{c}
\times SU(2)_L \times U(1)_Y$.

If  $F$  transforms as a 54 or 210 dimensional
representation of SO(10) \cite{Martin:2009ad}, one obtains the following relation
among the MSSM gaugino masses at $ M_{\rm GUT} $:
\begin{align}
M_3: M_2:M_1= 2:-3:-1 ,
\label{gaugino10}
\end{align}
where $M_1, M_2, M_3$ denote the gaugino masses  corresponding
to $U(1)$, $SU(2)_L$ and $SU(3)_c$, respectively. In order to obtain the
correct sign for the desired contribution to $(g-2)_{\mu}$, we choose
$M_{1} > 0 $, $M_{2} > 0$ and $M_{3} < 0$.
{Notice that, in  general, if $F^{ab}$ transforms {non trivially} under {SO(10)}, the
SSB terms such as the trilinear couplings and scalar mass terms are not
necessarily universal at $M_{GUT}$}. However, we can assume, consistent
with SO(10) gauge symmetry, that the coefficients associated with terms
that violate the SO(10)-invariant form are suitably small, except for
the gaugino term in Eq.(\ref{eq4}). {We also assume that
D-term contributions to the SSB {terms} are much smaller compared
with contributions from fields with non-zero auxiliary $F$-terms.}

Employing the boundary condition from Eq.(\ref{gaugino10}), one can define the MSSM gaugino
masses at $ M_{\rm GUT} $ in terms of the mass parameter $M_{1/2}$ :
\begin{align}
M_1&= M_{1/2} \nonumber \\
M_2&= 3M_{1/2} \nonumber \\
M_3&= - 2 M_{1/2}.
 \label{gaugino11}
\end{align}
Note that $M_2$ and $M_3$ have opposite signs which is important {in} implementing Yukawa coupling
unification as  shown in~\cite{Gogoladze:2009ug}.
In order to quantify Yukawa coupling unification,  following
\cite{Baer:2008jn},  we define the quantity $R_{tb\tau}$ as,
\begin{align}
R_{tb\tau}=\frac{ {\rm max}(Y_t,Y_b,Y_{\tau})} { {\rm min} (Y_t,Y_b,Y_{\tau})}.
\end{align}

We have performed random scans  in  the fundamental parameter space as follows:
\begin{align}
0\leq  m_{16}  \leq 10\, {\rm TeV} \nonumber \\
0\leq  M_{1/2}  \leq 5\, {\rm TeV} \nonumber \\
0\leq  m_{10}  \leq 10\, {\rm TeV} \nonumber \\
-3\leq A_{0}/m_{16} \leq 2 \, {\rm TeV} \nonumber \\
1.1\leq \tan\beta \leq 60\, \nonumber \\
0\leq  m_{t}  \leq 220\, {\rm GeV} \nonumber \\
\mu>0.
\label{parameterRange-NUHM1}
\end{align}
Here $ m_{16} $ is the universal SSB mass for MSSM sfermions, $ m_{10} $ is the universal
SSB mass term for the  up/down  MSSM Higgs doublets, $ M_{1/2} $ is the gaugino mass parameter,
tan$\beta $ is the ratio of the vacuum expectation values (VEVs) of the two MSSM Higgs doublets,
$ A_{0} $ is the universal SSB trilinear scalar interaction (with corresponding Yukawa couplings factored out),
$m_t$ denotes the top-quark mass and $\mu>0$ sets the sign for the bi-linear SSB Higgs mixing term
whose absolute value is fixed by requiring REWSB.

\begin{figure}[t!]
\subfigure{\includegraphics[scale=.9]{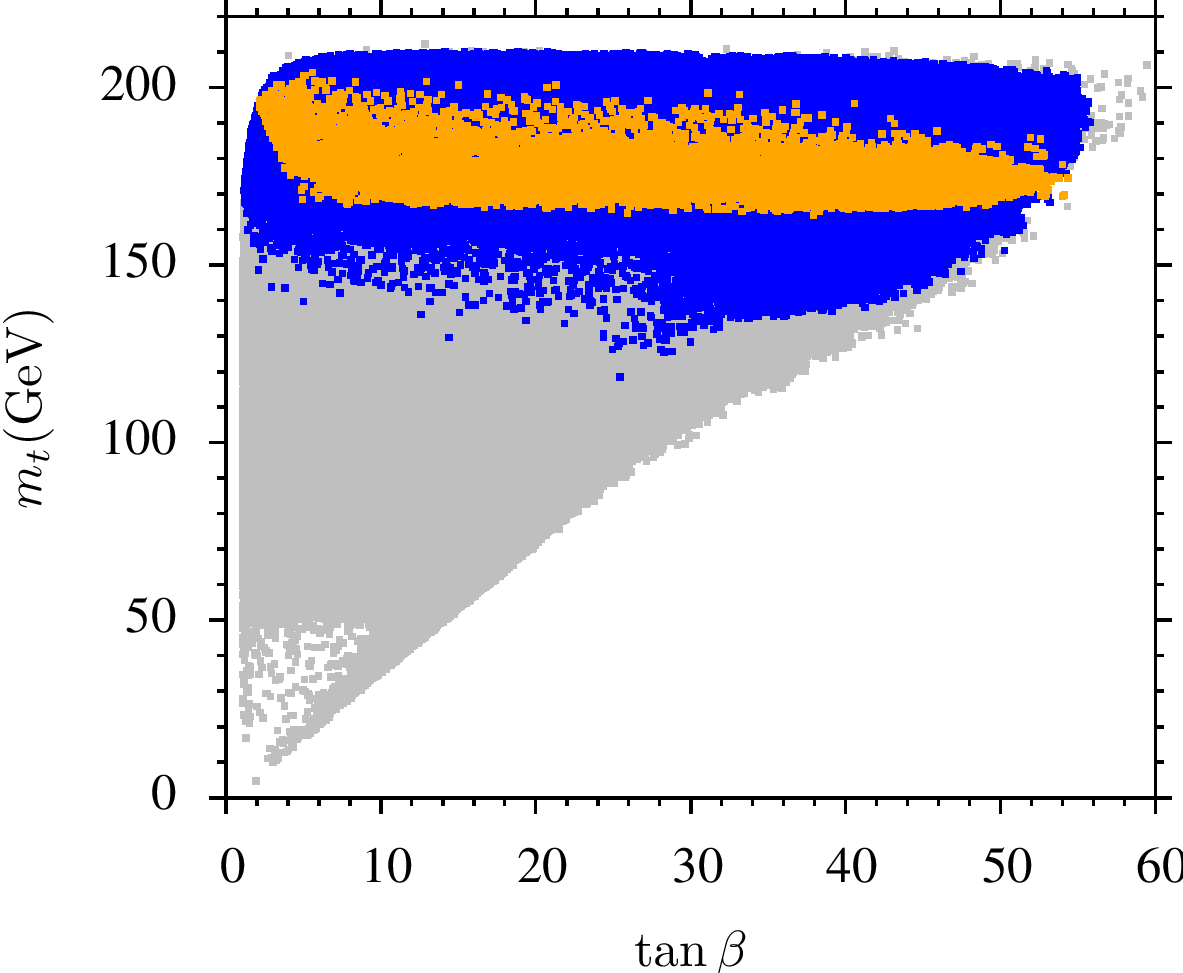}}
\subfigure{\includegraphics[scale=0.9]{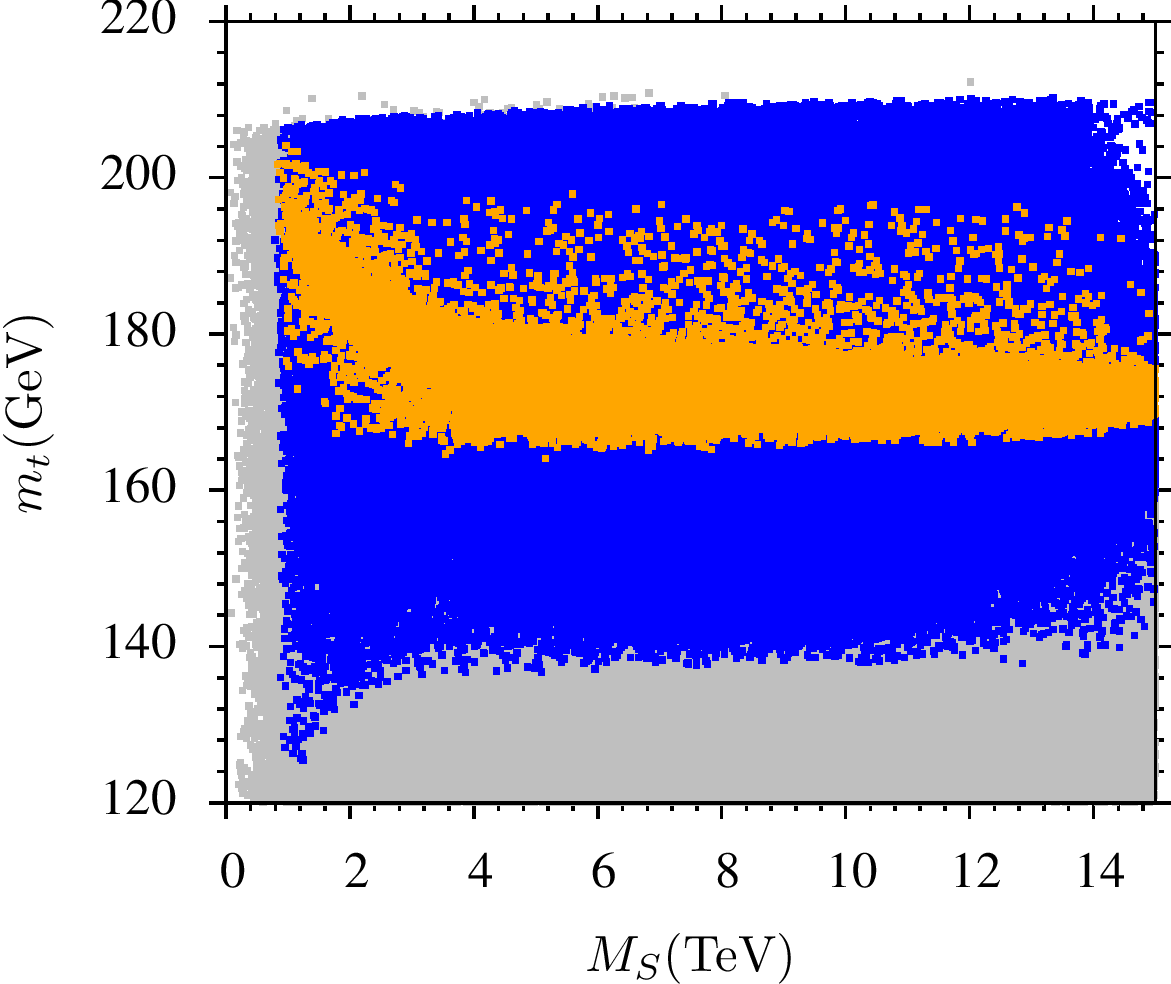}}
\caption{Plots in  the $m_t$-tan$\beta$  and $m_t$-$M_S$ planes.
{\it Gray} points are consistent with REWSB  and neutralino LSP.  {\it Blue} points form a subset of the {\it gray}
and satisfy sparticle mass bounds and other constraints described in Section \ref{pheno}. {\it Orange} points
belong to a subset of {\it blue} points and satisfy the lightest CP even Higgs boson
mass bound $123~{\rm GeV}\leq m_h\leq 127~{\rm GeV}$.
\label{figure1}}
\end{figure}

In Figure~\ref{figure1}, we present our results in the $m_t$-tan$\beta$  and $m_t$-$M_S$ planes.
{\it Gray} points are consistent with REWSB  and neutralino LSP.  {\it Blue} points form a subset of the {\it gray} ones
and satisfy sparticle mass bounds and other constraints described in Section \ref{pheno}.
{\it Orange} points belong to a subset of {\it blue} points and satisfy the lightest CP even Higgs
boson mass bound $123~{\rm GeV}\leq m_h\leq 127~{\rm GeV}$.

It is interesting to note that REWSB can arise even  for  $m_t$ values as low as 10 GeV, but
tan$\beta$ values are then constrained  to be in a narrow range.
For a given top quark mass we have a well defined tan$\beta$ interval from  the
requirement of REWSB. For instance,  from the $m_t$-tan$\beta$ plane, one sees that  in order to have
REWSB with  $m_t\approx 10$ GeV, the value of tan$\beta$ should  be around 3 or so. Applying all the collider and
B-physics constraints except the lightest Higgs boson mass bound, we obtain  for the
top quark mass the bound  $125~{\rm GeV}\lesssim m_t\lesssim 208~{\rm GeV}$ ({\it blue} points in Figure~\ref{figure1}).
 Applying next the   Higgs boson mass bound $123~{\rm GeV}\leq m_h \leq 127~{\rm GeV}$,
 the top quark is expected  lie  in the interval $164~{\rm GeV}\lesssim m_t\lesssim 205~{\rm GeV}$.
 Similar observations can be  made from

 the $m_t$-$M_S$ plane. It is evident from this plane that no matter how heavy the stop  mass,
 the top quark cannot  be lighter than 164 GeV  in low scale supersymmetry when light CP-even Higgs boson is in this
 the interval
  $123~{\rm GeV}\leq m_h \leq 127~{\rm GeV}$. This lower bound is very close  to the values
  obtained from fits to the  electroweak data \cite{Baak:2012kk,Eberhardt:2012gv}.

\begin{figure}[t!]

\subfigure{\includegraphics[scale=.9]{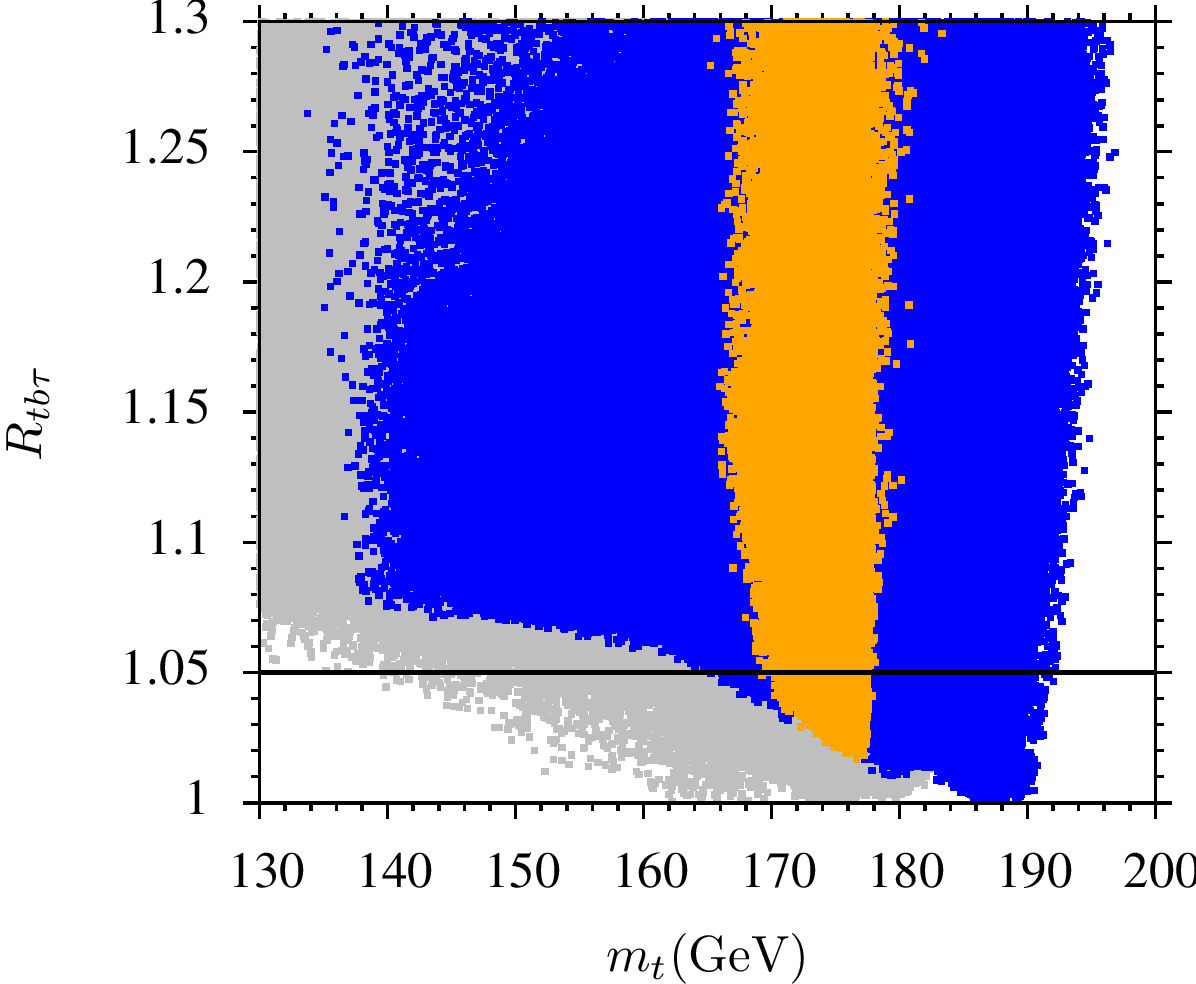}}
\subfigure{\includegraphics[scale=0.9]{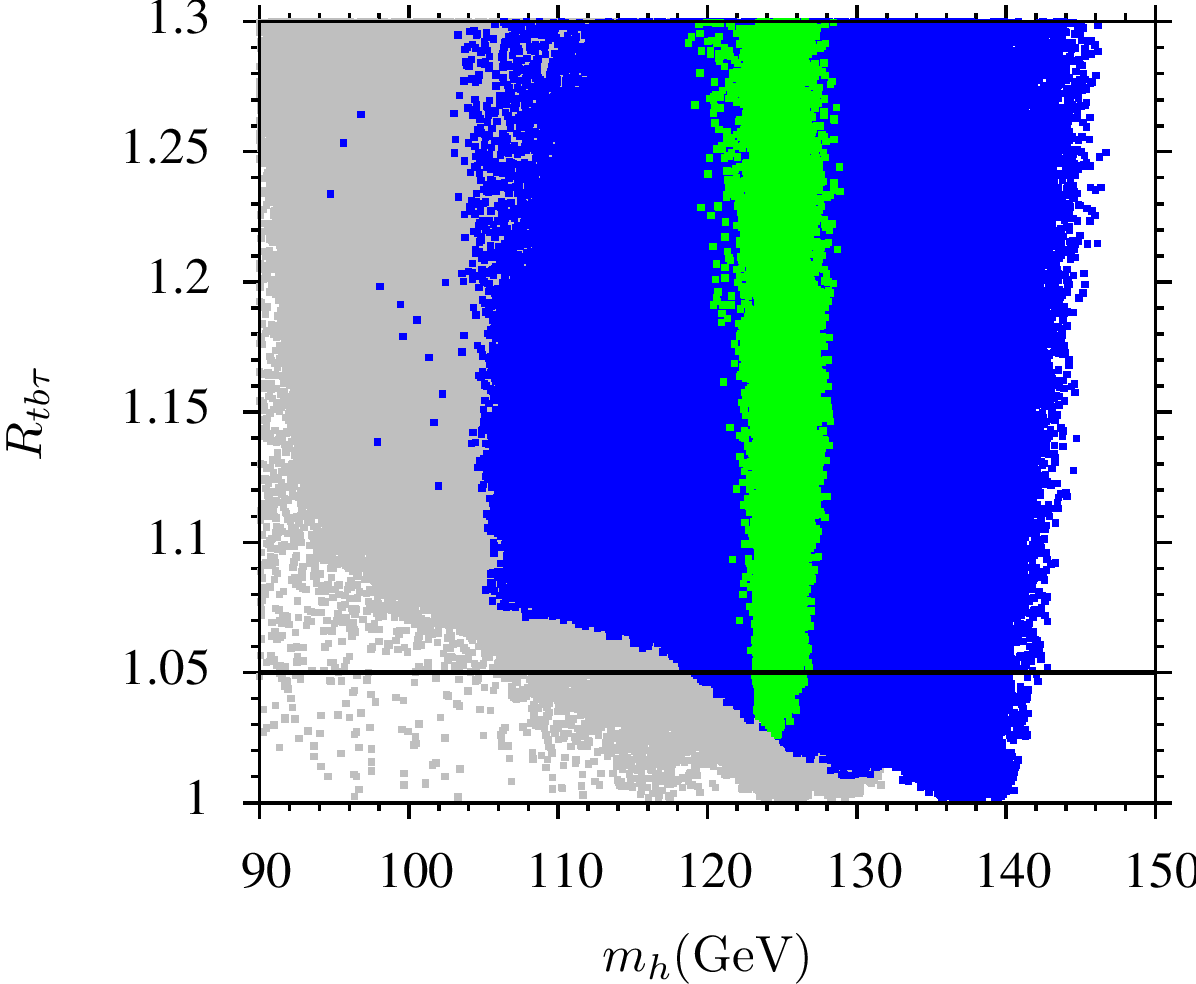}}
\caption{Plots in the $R_{tb\tau}$-$m_t$ and $R_{tb\tau}$-$m_h$ planes.
 Color coding is  same as in  Figure~\ref{figure1}. The {\it green} points in the $R_{tb\tau}$ - $m_h$ plane
form a   subset of {\it blue} points and satisfy    the
 bound $172.3~{\rm GeV}\leq m_t\leq 174.1~{\rm GeV}$.
\label{figure2}}
\end{figure}

In Figure~\ref{figure2} we present  the results in the $R_{tb\tau}$-$m_t$ and $R_{tb\tau}$-$m_h$ planes.
The color coding is the same as in  Figure~\ref{figure1}. The {\it green}  points in the $R_{tb\tau}$ - $m_h$ plane
form a   subset of {\it blue} points and satisfy    the
 bound $172.3~{\rm GeV}\leq m_t\leq 174.1~{\rm GeV}$.

The horizontal lines in the two planes shown in Figure~\ref{figure2} correspond to
5\% $t$-$b$-$\tau$ Yukawa coupling unification ($R_{tb\tau}=1.05$).
From the  $R_{tb\tau}$-$m_t$ plane, we see that without
requiring any constraints other than REWSB,
$t$-$b$-$\tau$ Yukawa unification  better than $5\%$ predicts that  the top quark cannot be lighter than 140 GeV.   The top quark mass
 is further constrained by
imposing the phenomenological constraints mentioned in Section~\ref{pheno}. Even without accounting for  the
bound on the light CP even Higgs mass ({\it blue} points), the top quark mass is found to lie
 in  the range $166~{\rm GeV}< m_t<192~{\rm GeV}$.
 While it is not evident from  Figure~\ref{figure2},
{from} {of}  the constraints on the top quark mass implemented in {\it blue}, the most severe  comes
from  observation of the decay $B_s\rightarrow \mu^+ \mu^-$. It is well known that
in low scale supersymmetry models, this flavor-changing decay receives contributions from the exchange
of the pseudoscalar Higgs boson $A$ \cite{Choudhury:1998ze},  and its branching ratio is
proportional to (tan$\beta)^6/ m_A^4$. Since $t$-$b$-$\tau$ Yukawa unification happens
for  large tan$\beta (\thickapprox 47$)
and it prefers relatively small values for the CP  odd pseudoscalar mass ($m_A< 2~{\rm TeV}$)
in this model \cite{Gogoladze:2011aa}, the top quark mass is severely constrained
by the $B_s\rightarrow \mu^+ \mu^-$ decay.
Applying the light CP even Higgs mass bound, the top quark is predicted to have a mass  in the narrow window $170~{\rm GeV}  \leq m_t \leq 178~{\rm GeV}$.

Next let us consider the constraints on the Higgs boson mass.
From the $R_{tb\tau}$-$m_h$ plane, taking into account  the collider and B-physics bounds
 (but excluding the  top quark mass bound),
$t$-$b$-$\tau$ YU better than $5\%$ requires
that the  light CP-even Higgs boson mass $m_h> 119$ GeV.  After imposing the
$1\sigma$  top quark mass bound, the model  predicts that the Higgs mass lies  in the
range  $122~{\rm GeV}\leq m_h \leq 126~{\rm GeV}$, in good agreement  with the current experimental observations.

\begin{figure}[t!]
\centering
\includegraphics[scale=0.9]{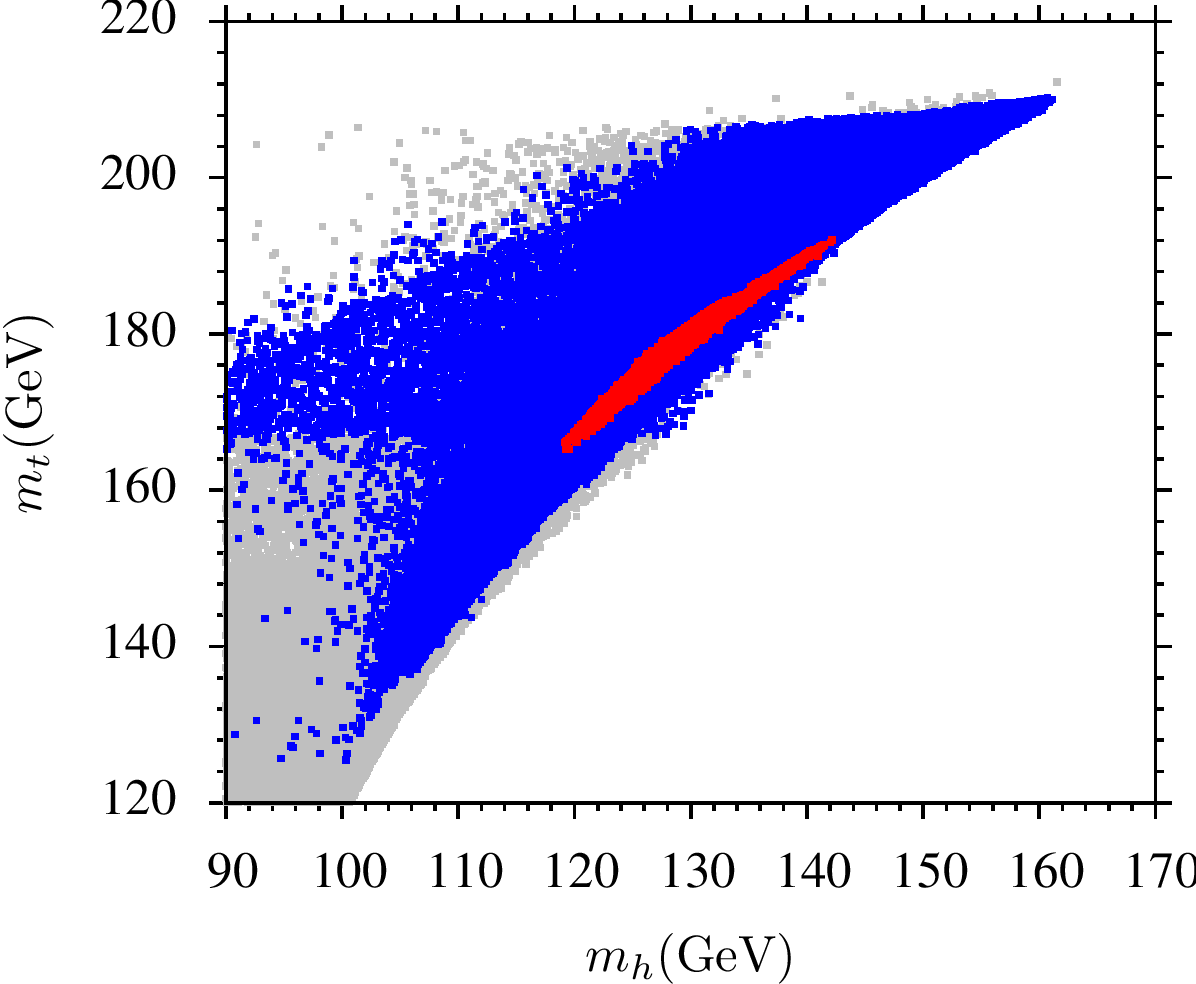}
\includegraphics[scale=0.9]{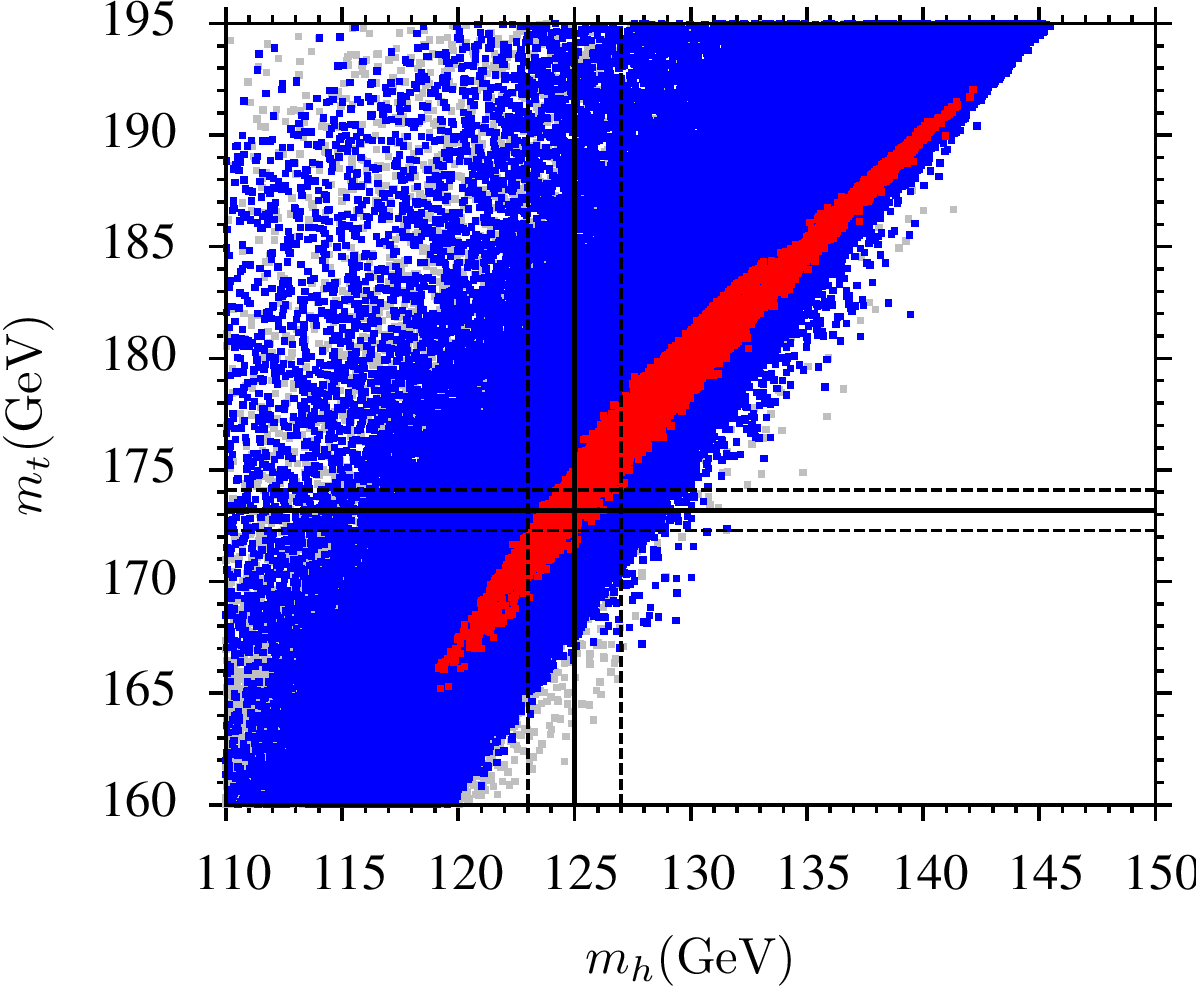}
\caption{Plots in the $m_t$-$m_h$ plane. The right panel is just a zoomed-in
version of the left panel. The color coding is
the same as in Figure~\ref{figure1} with the addition of {\it red} points which
form a subset of {\it blue} points and satisfy  $R_{tb\tau}< 1.05$.\label{figure3}}
\end{figure}

We display  the correlation between the top quark and Higgs boson masses  in the presence of Yukawa unification,
   in the $m_t$ - $m_h$ plane  in Figure~\ref{figure3}.
The right panel is  a zoomed-in version of the left panel in this figure. The color coding is
the same as in Figure~\ref{figure1},  with the addition of {\it red} points  which
form a subset of { \it blue} points and satisfy  $R_{tb\tau}< 1.05$.

The sharp edge towards the right in  the $m_t$-$m_h$ plane
shows that one requires  need a heavy top quark in order to obtain a Higgs boson mass
of 125 GeV. We also observe  that requiring $5\%$ or better Yukawa unification  makes
the Higgs and top quark masses   strongly correlated.
Requiring Yukawa unification along with  a 125 GeV Higgs boson mass,
one predicts the top quark mass to be in  the interval $172~{\rm GeV}\leq m_t \leq 175~{\rm GeV}$.
 Conversely, {requiring} Yukawa unification along with the top quark
mass to be at its experimentally observed central value, the Higgs boson mass is
predicted to be in the {range $124~{\rm GeV}\leq  m_h \leq 126~{\rm GeV}$}.

While there is a few GeV theoretical error in
the  calculation of the
Higgs mass \cite{Draper:2013oza}, the strong correlation between the Higgs boson
mass, the top quark mass and Yukawa unification seems to be quite compelling   { in this class of
models.}

\section{SO(10) GUT with  universal gauginos masses}

In this section we consider $t$-$b$-$\tau$ Yukawa unification in a supersymmetric
SO(10) GUT model with universal SSB gauginos masses and ``Just so'' Higgs mass splitting \cite{Blazek:2002ta}.
This GUT scale boundary condition is commonly known as the non-universal Higgs mass  (NUHM2) model.
The SSB parameters include  $m_{16},~ M_{1/2},~ m_{H_d},~ m_{H_u},~ A_0,~ {\rm \tan}\beta$.
This model  predicts a  heavy sfermion
spectrum ($m_{16}\gtrsim 20$ TeV) but relatively light gaugino masses~\cite{Baer:2008jn}.
For instance, the gluinos  in this scenario are not heavier than 3 TeV or so \cite{Baer:2012cp},
which can be tested at the LHC.  We next  investigate the allowed range   for the top quark and
Higgs boson masses   in this scenario.

We have performed random scans for the following ranges of parameters:

\begin{align}
0\leq  m_{16}  \leq 21\, {\rm TeV} \nonumber \\
0\leq  M_{1/2}  \leq 5\, {\rm TeV} \nonumber \\
0\leq  m_{H_d}  \leq 27\, {\rm TeV} \nonumber \\
0\leq  m_{H_u}  \leq 25\, {\rm TeV} \nonumber \\
-3\leq A_{0}/m_{16} \leq 3 \, {\rm TeV} \nonumber \\
1.1\leq \tan\beta \leq 60\,  \nonumber \\
0\leq  m_{t}  \leq 220\, {\rm GeV} \nonumber\\
\mu> 0
\label{parameterRange-NUHM2}
\end{align}

\begin{figure}[t!]
\subfigure{\includegraphics[scale=.9]{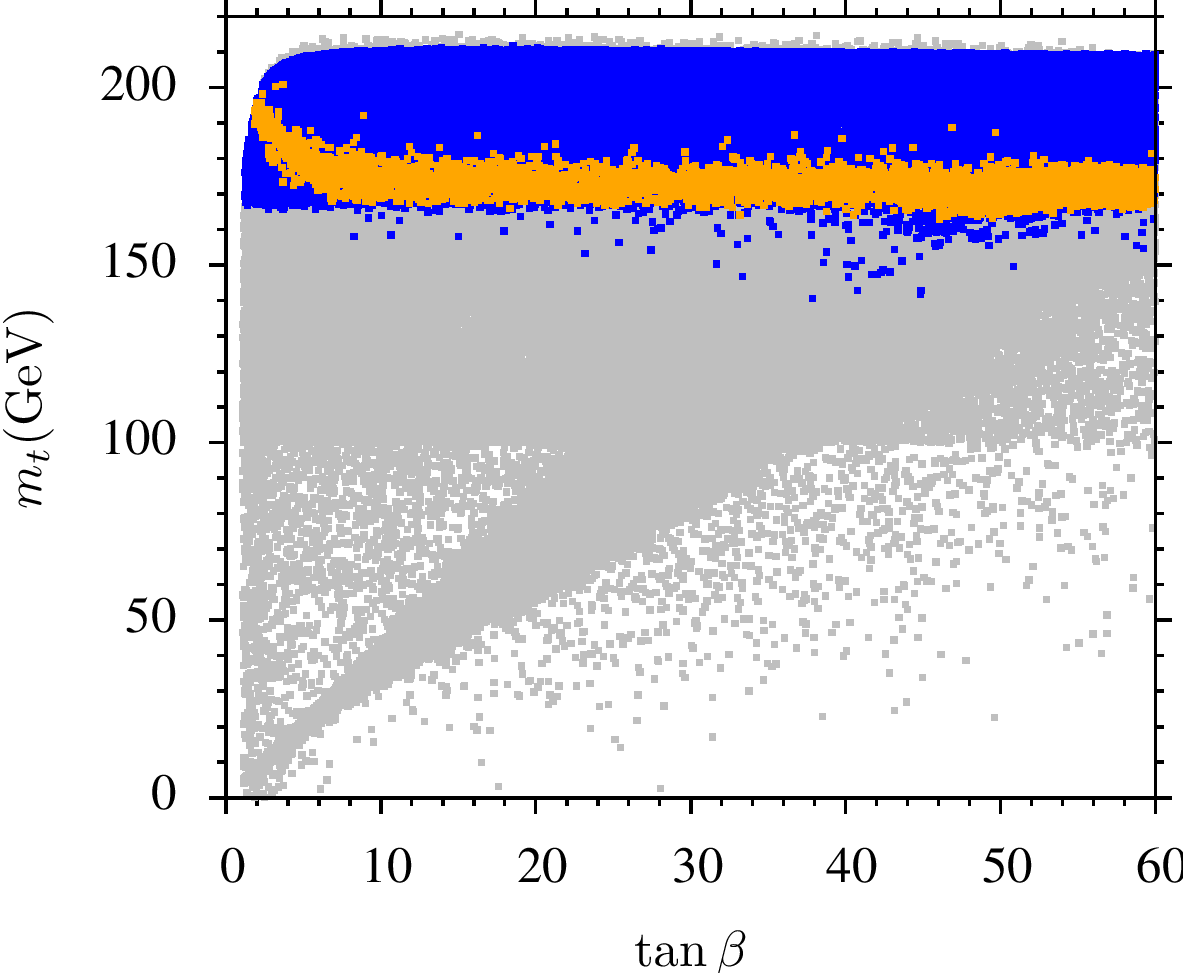}}
\subfigure{\includegraphics[scale=0.9]{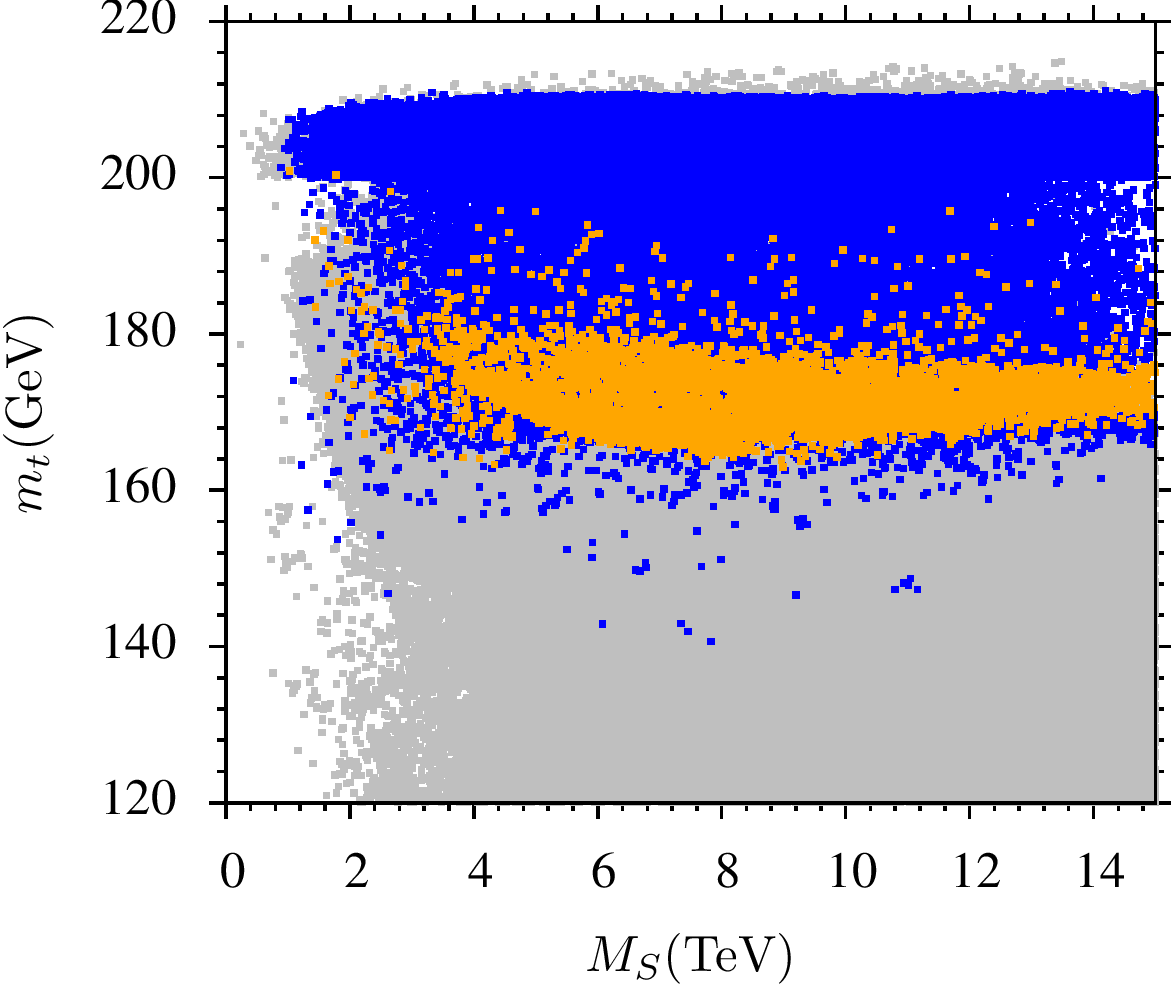}}
\caption{Plots in the $m_t$-tan$\beta$ and $m_t$-$M_S$ planes.  Color coding same as in
Figure~\ref{figure1}.
\label{figure4}}
\end{figure}
In Figure~\ref{figure4}, we present results in the $m_t$-tan$\beta$ and $m_t$-$M_S$ planes.
 The color coding is the same as in Figure~\ref{figure1}

In contrast  to the SO(10) model discussed in Section 3, { one} can
see from the $m_t$ - tan$\beta$ plane that   there is no restriction
 in this version of SO(10) on either the top quark mass or the value of tan$\beta$  from the point of view of
REWSB.
After applying sparticle mass bounds and constraints  from B-physics,
 the lower bound on the top quark mass
is dramatically  increased ({\it blue} points in Figure~\ref{figure4}). The lower bound
for {\it blue} points in Figure~\ref{figure4}  appears to be higher compared
to the bounds obtained from Figure~\ref{figure1}.
However, this is likely
due to a lack of statistics, as indicated in  the $m_t$ - $M_S$ plane,
Figure~\ref{figure4}. Indeed, we have collected far less
data  in this case compared to the model presented in Section 3.
A more exhaustive  study will likely fill the regions
 around the isolated {\it blue} points in $m_t$-tan$\beta$ plane.
The main reason for a lack of extensive  statistics for this model is that  we
 are mainly  interested in the correlation between the top quark and Higgs boson masses, and this
is amply illustrated by the data that we have collected as is obvious by focusing on the
{\it orange} points in Figure~\ref{figure4}. As a reminder to the reader, the {\it orange} points
form a subset of the {\it blue} ones and require the Higgs boson to have a mass in the
range $123~{\rm GeV}\leq m_h\leq 127$ GeV. This constraint
on $m_h$ requires, in turn, that $m_t$ is restricted to lie in the range
$164~{\rm GeV}< m_t<200~{\rm GeV}$.

\begin{figure}[b!]
\subfigure{\includegraphics[scale=.9]{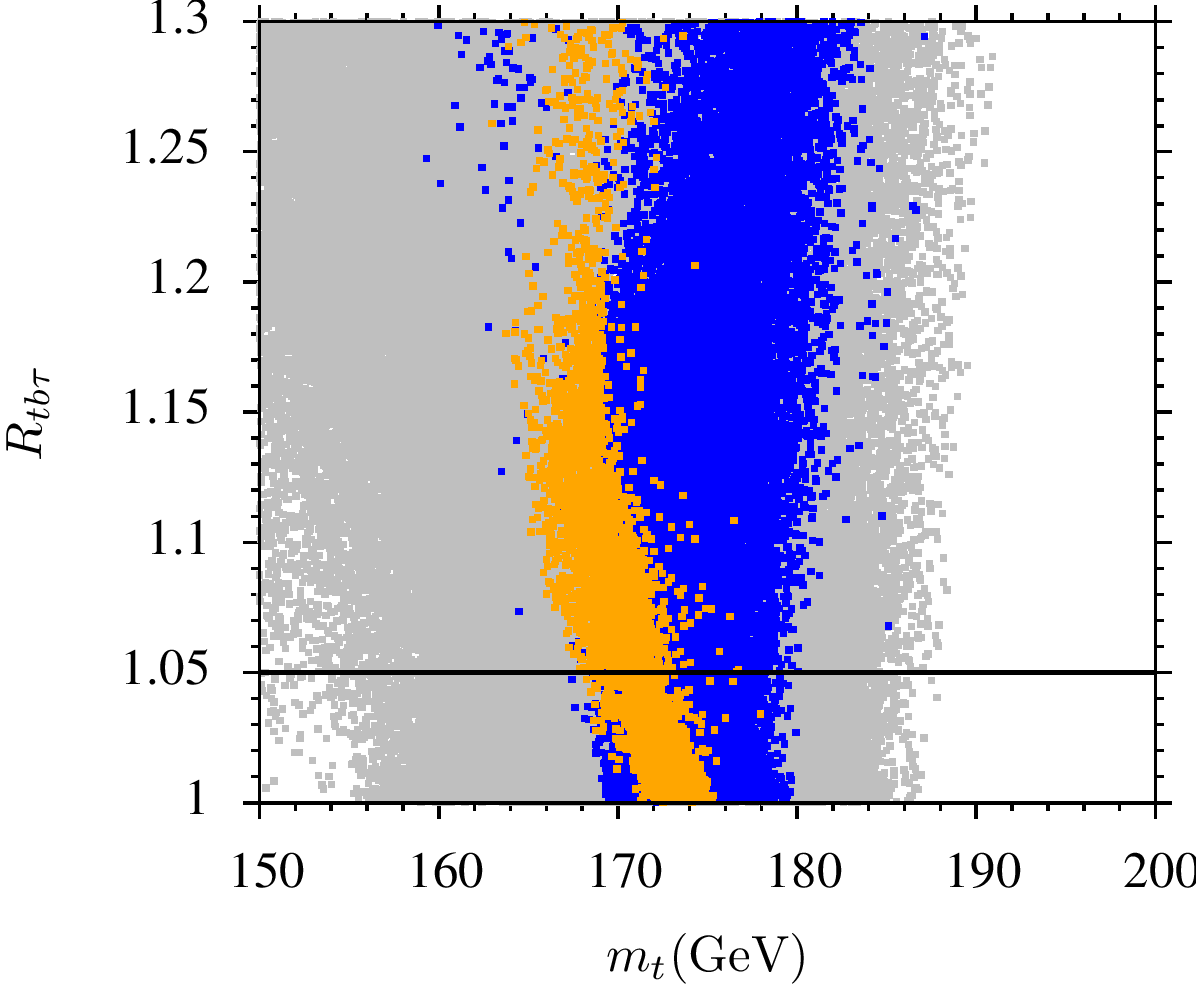}}
\subfigure{\includegraphics[scale=0.9]{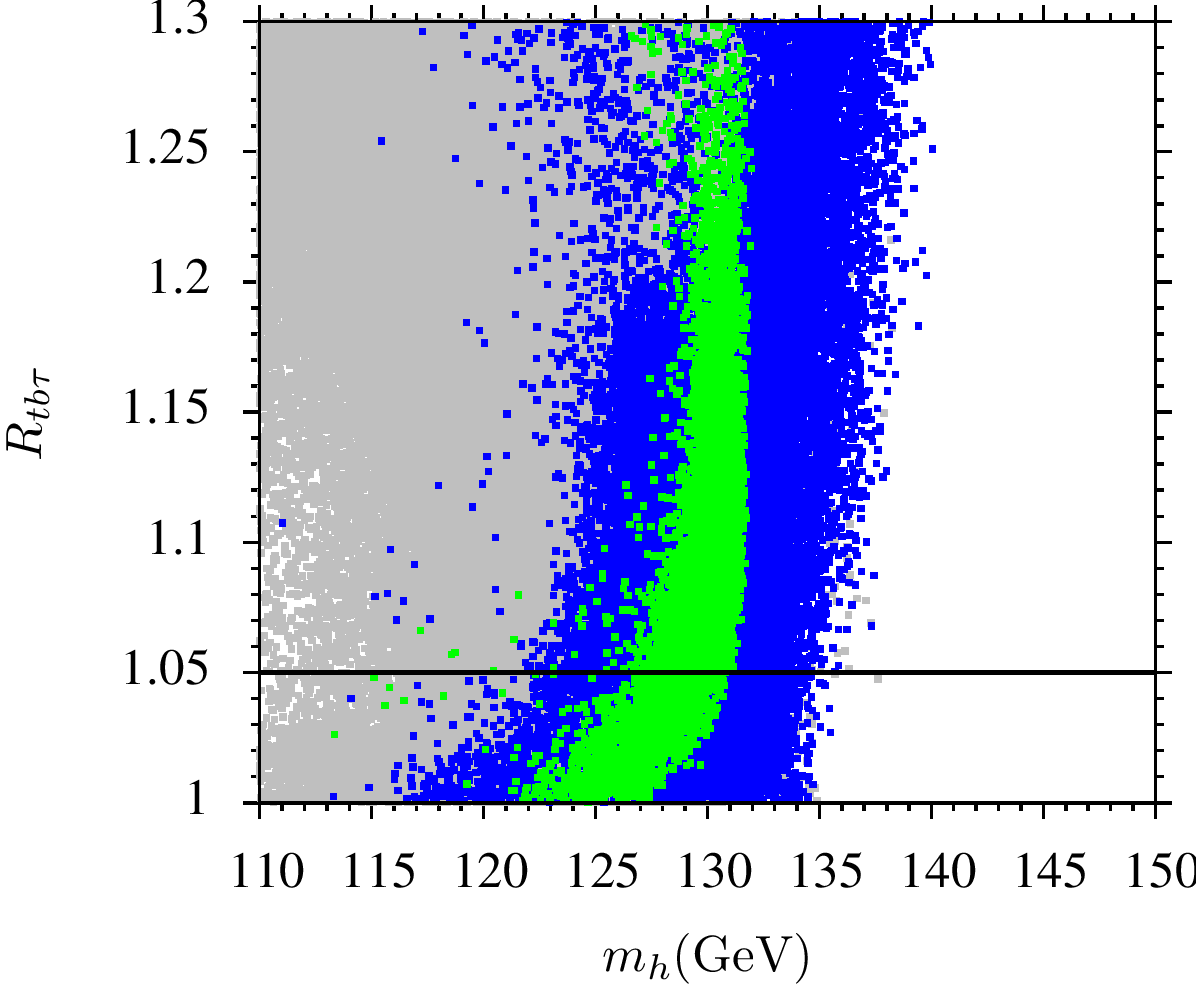}}
\caption{Plots in $R_{tb\tau}$- $m_t$ and $R_{tb\tau}$-$m_h$ planes. The color coding is the same as  Figure~\ref{figure2}.
{\it Green} points in the $R_{tb\tau}$-$m_t$ plane
belong to a subset of {\it blue} points and satisfy the top quark
mass bound $172.3~{\rm GeV}\leq m_t\leq 174.1~{\rm GeV}$.
\label{figure5}}
\end{figure}

In Figure~\ref{figure5} we show the results in the $R_{tb\tau}$-$m_t$ and $R_{tb\tau}$-$m_h$ planes.
The color coding is the same as in Figure 4. The {\it green} points in the $R_{tb\tau}$-$m_t$ plane
belong to a subset of {\it blue} points and satisfy the 1$\sigma$ top quark
mass bound $172.3~{\rm GeV}\leq m_t\leq 174.1~{\rm GeV}$.
%

As shown in Figure~\ref{figure5},
requiring REWSB  and $t$-$b$-$\tau$
Yukawa unification better than $5\%$ requires that $m_t$ is heavier than 140 GeV and
 lighter than 188 GeV.  The top quark mass   gets confined
to the range $166~{\rm GeV}< m_t< 180~{\rm GeV}$ after applying the
collider and B-physics constraints on the data from  Section~\ref{pheno}
(excluding the Higgs boson mass bound).
This bound is  virtually unchanged after applying
the experimental bound on the Higgs mass  ({\it orange} points in the figure).
The most  severe constraint which gives rise to  the {\it blue} region comes from the decay  $b\rightarrow s\gamma$.
This is because in low scale  supersymmetry the dominant contribution to  {the} $b\rightarrow s\gamma$ branching ratio
 is proportional to $\mu A_t{\rm \tan}\beta$, {while
Yukawa unification with universal gaugino masses at $M_{\rm GUT}$ requires $A_t\gtrsim 15$ TeV \cite{Baer:2008jn}}.
The SUSY contribution to  $BR(B_s\rightarrow \mu^+ \mu^-)$
is suppressed in this case because, as pointed out in~\cite{Baer:2012cp}, $m_A>3$ TeV
in this scenario.

\begin{figure}[b!]

\subfigure{\includegraphics[scale=.9]{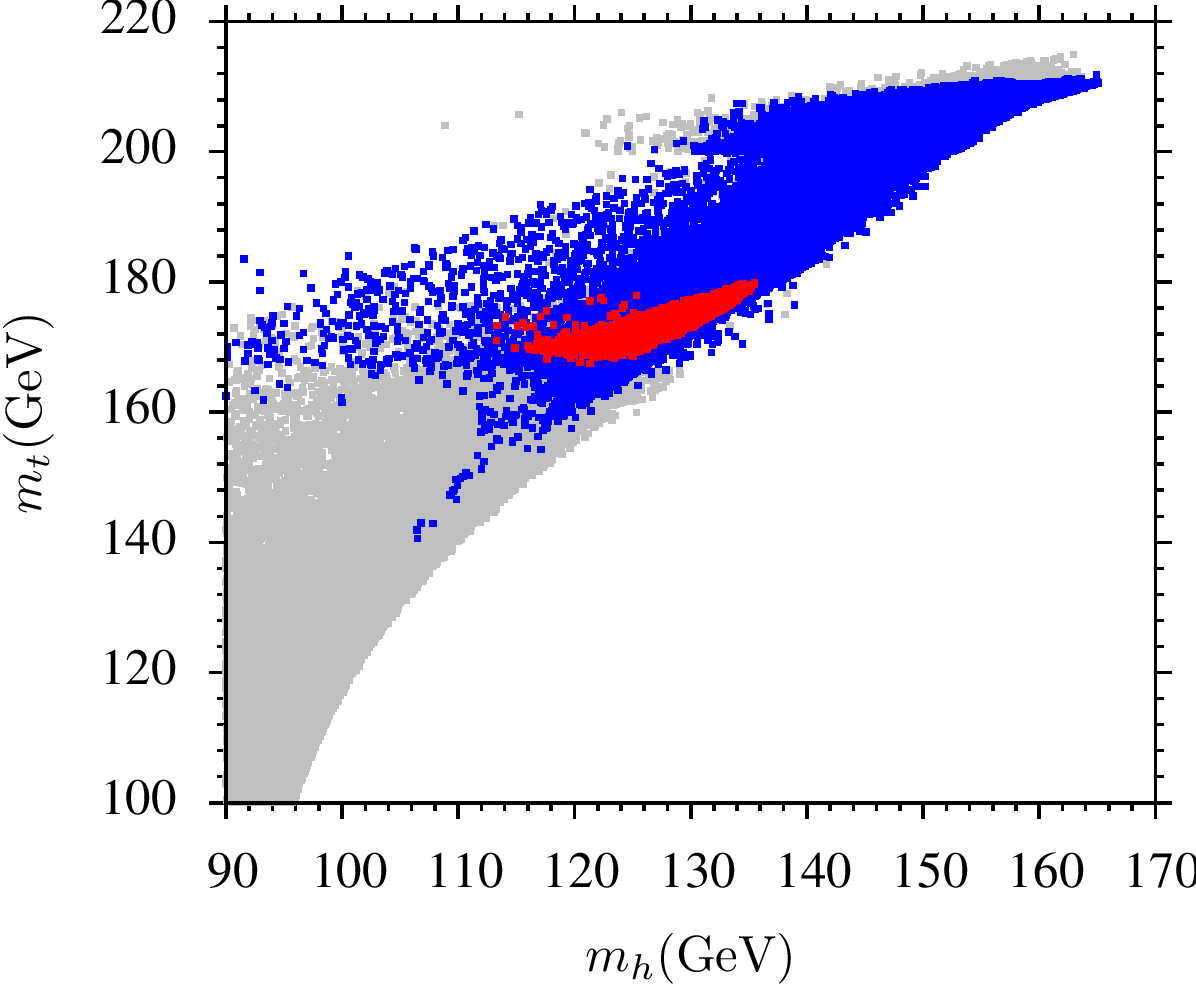}}
\subfigure{\includegraphics[scale=.9]{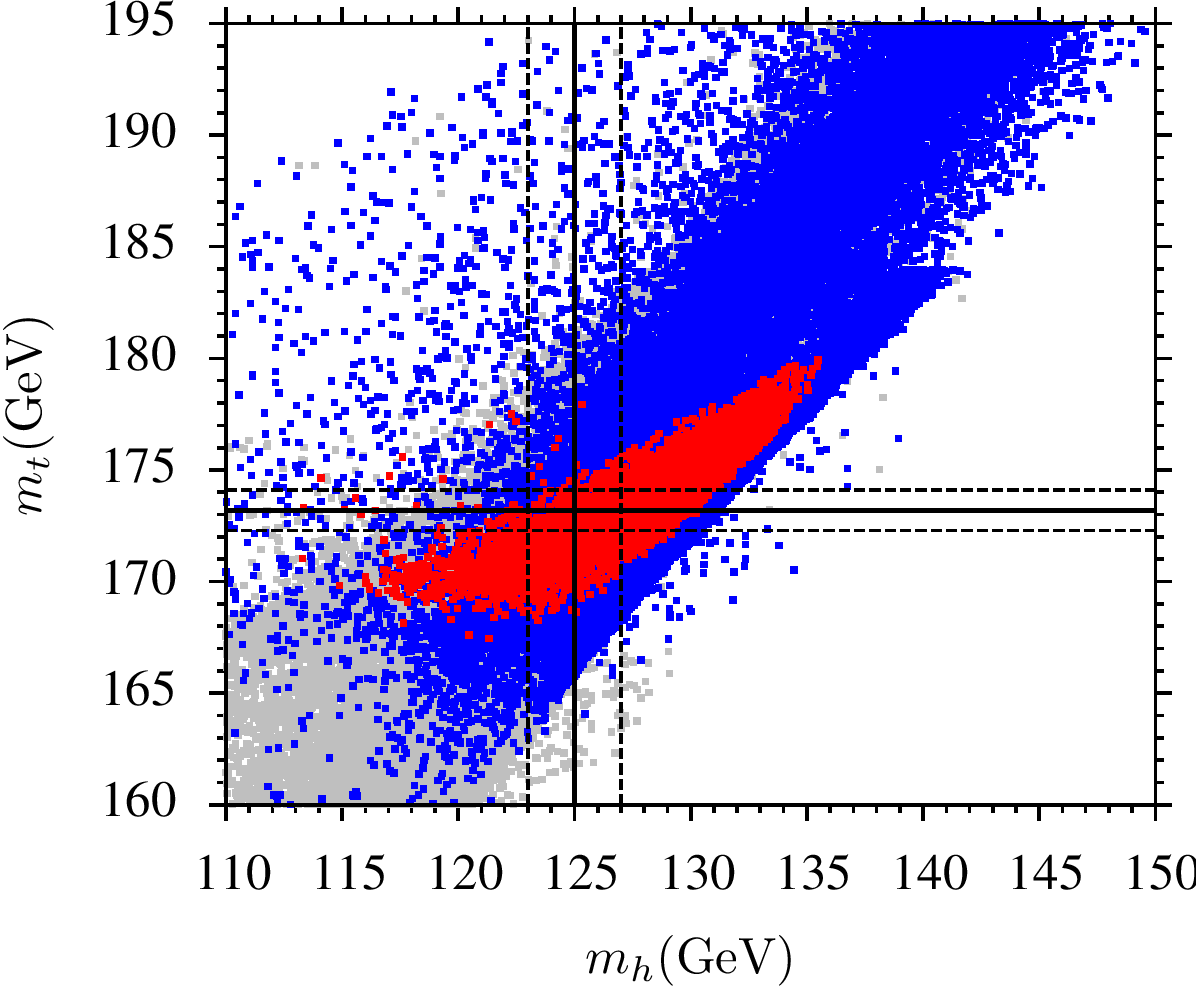}}
\caption{ Plots in $m_t$-$m_h$ plane. Color coding same as  Figure~\ref{figure3}.
\label{figure6}}
\end{figure}

The results in the $m_t$-$m_h$ plane  in Figure~\ref{figure6} show the correlation between the top quark and Higgs boson masses
 in the presence  of  Yukawa unification. This figure is similar in spirit to Figure~\ref{figure3} and shares its color coding.

The sharp right edge in the $m_t$-$m_h$ plane is retained in this version of the SO(10) model,
and it  shows that one  needs a heavy top quark in
order to obtain a 125 GeV Higgs boson. We also note  that $5\%$ or better Yukawa unification
(red points)  makes the Higgs and top quark mass interdependence   stronger  as in the SO(10) model of  Section 3.
Requiring the Higgs boson to have a mass of 125 GeV yields   a top quark mass in the interval
$168~{\rm GeV} \leq m_t \leq 177~{\rm GeV}$.
However,  fixing the top quark mass in this case does not yield  a sharp prediction for
the Higgs boson mass.
For instance, for the measured  central value of $m_t$,  the Higgs boson mass lies in  the
range $113~{\rm GeV} \leq m_h \leq 131~{\rm GeV}$.

\begin{figure}[t!]
\centering
\includegraphics[scale=0.9]{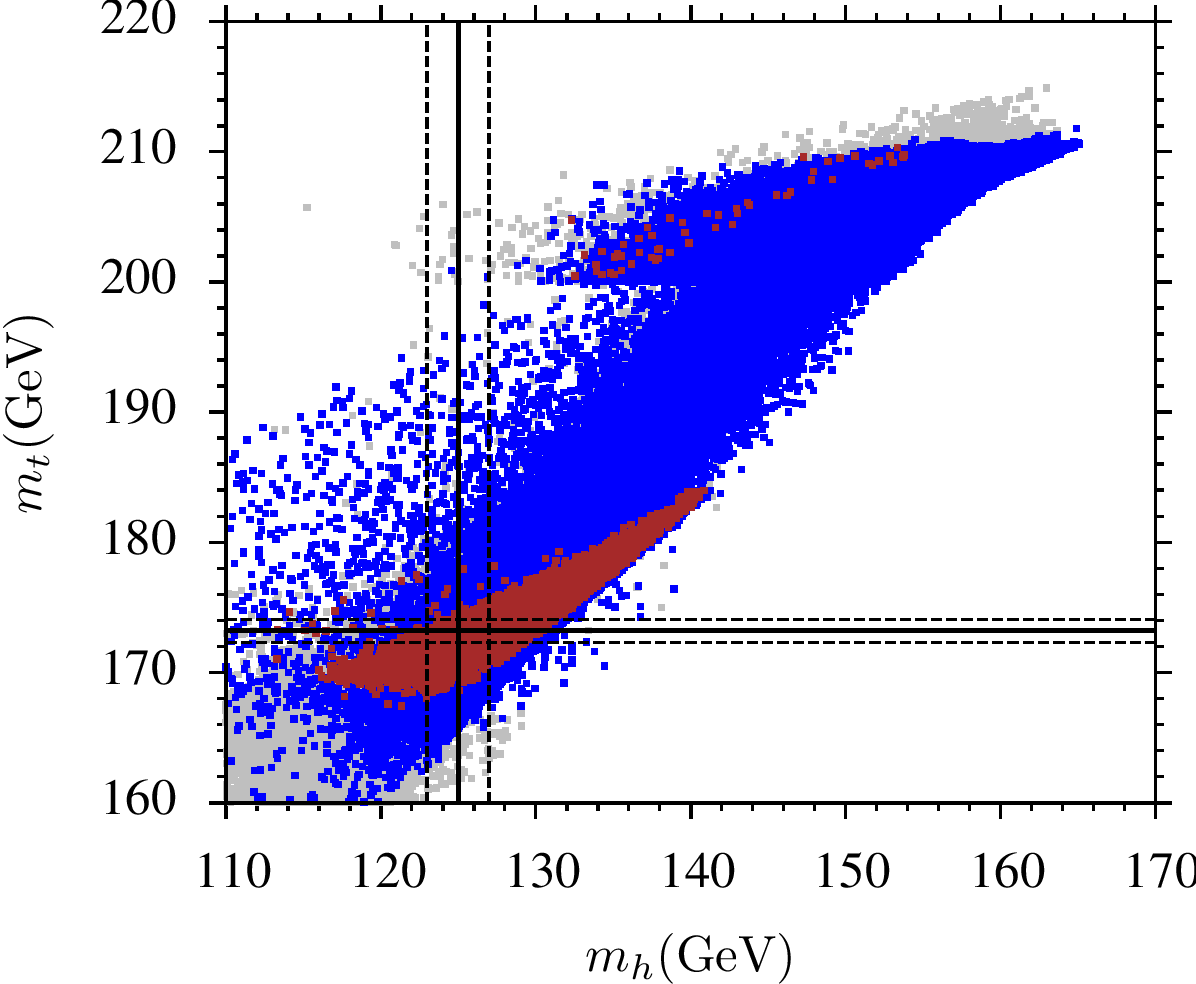}
\caption{ Plots in $m_t$-$m_h$ plane. The color coding is the same as  Figure~\ref{figure1}.
The brown points form a subset of {\it blue} points and represent 5\% or better $b$-$\tau$ Yukawa
unification.
\label{figure7}
}
\end{figure}

 It is interesting to consider $b$-$\tau$ Yukawa unification
which  may be more natural with  non-universal SSB mass terms
 in the Higgs  sector.
 In Figure~\ref{figure7} we present
results in the $m_t$-$m_h$ plane with brown points
 signifying $5\%$ or better $b$-$\tau$ Yukawa unification. The rest
of the color coding is the same as in Figure~\ref{figure1}.
We can see two distinct brown regions corresponding to $b$-$\tau$ Yukawa unification
in this figure. The island-like region is nothing but a somewhat larger
region than the {\it red} region of Figure~\ref{figure6}. It is larger   because
$b$-$\tau$ Yukawa unification is  less restrictive than $t$-$b$-$\tau$
Yukawa unification. Unlike $t$-$b$-$\tau$ Yukawa unification, $b$-$\tau$ Yukawa unification
is realized  for  both large and small tan$\beta$ values. The region formed by the isolated
brown points in Figure~\ref{figure7} corresponds to the case of small $\tan\beta$.
As pointed out
in~\cite{Gogoladze:2011be},  $b$-$\tau$ Yukawa unification for  small
tan$\beta$ values  requires that the  sfermion SSB mass terms are larger than 5 TeV or so.
Moreover, the  non-universal SSB sfermion  mass {term} {terms} at $M_{\rm GUT}$
{are} { need to be} based on their SU(5) representation.
While we do not have non-universal SSB mass terms characteristic of $SU(5)$, the top quark mass
is free in our analysis. As a result,  $b$-$\tau$ Yukawa
unification  does occur for  small tan$\beta$ values,
but it occurs in an experimentally unacceptable region for the top quark
mass, namely $200$-$210$ GeV.

\section{Conclusions}

We have tried to understand the top quark mass from the
requirement that $t$-$b$-$\tau$  Yukawa coupling unification occurs at $M_{\rm GUT}$.
For this, we consider two $SO(10)$ GUT  models, with  one model  having universal SSB gaugino masses   but
non-universal SSB Higgs  mass terms. In the second example we have   universal Higgs SSB mass terms
 but
non-universal SSB gaugino masses. We have also considered  the correlation between the Higgs boson and
top quark masses.
The upper bound  $m_t \sim 210$ GeV or so on the top quark mass comes from requiring
perturbativity of the top quark Yukawa coupling up to  scale $M_{\rm GUT}$. Radiative
electroweak symmetry breaking imposes a lower bound on the top quark mass, which depends on $\tan\beta$. In the model with non-universal  SSB
gaugino  masses, applying all the collider and B-physics constraints
including the bound on the Higgs boson mass yields  the interval
$164~{\rm GeV}\lesssim m_t \lesssim 205~{\rm GeV}$.  Further, imposing
Yukawa coupling unification narrows  this
mass range,  namely  $172~{\rm GeV}\lesssim m_t \lesssim 175~{\rm GeV}$.

For the model with non-universal SSB Higgs  mass terms, imposing the collider
bounds including the bound from   the Higgs boson mass, one arrives at a similar
range for the top quark mass as the previously quoted result. However, the allowed top
quark mass range after imposing Yukawa coupling unification is somewhat more
relaxed than in the previous case, to wit  $168~{\rm GeV}\lesssim m_t \lesssim 177~{\rm GeV}$, which  is
quite consistent with the experimental data. Furthermore requiring $b$-$\tau$
 instead of $t$-$b$-$\tau$ Yukawa unification in this model relaxes the top quark range,
 $167~{\rm GeV}\lesssim m_t \lesssim 182~{\rm GeV}$, for large tan$\beta$ values.
 The small tan$\beta$ scenario in this model predicts a top quark mass
in the neighborhood of $200$ GeV or so, and is, therefore, disfavored.

The correlation between the Higgs boson mass and the top quark mass is also very
interesting, particularly in the model with non-universal SSB gaugino  mass terms. By
imposing $t$-$b$-$\tau$ Yukawa coupling unification and requiring the top quark mass to be
close to its observed central value,  the Higgs boson mass is founded to lie  in the {range
$124~{\rm GeV}\lesssim m_h \lesssim 126~{\rm GeV}$}. This correlation between $m_t$ and $m_h$ from Yukawa unification   is not as strong in the
model with non-universal SSB Higgs  mass terms. In this case we obtain the result  $113~{\rm GeV}\lesssim m_h \lesssim 131~{\rm GeV}$.

\section*{Acknowledgments}
We would like to thank K.S. Babu, David Shih, Ayres Freitas, Wang Kai
and Tariq Saeed for useful discussions.
This work is supported in part by the DOE Grant No. DE-FG02-12ER41808
This work used the Extreme Science and Engineering Discovery Environment (XSEDE), which is
supported by the National Science Foundation grant number OCI-1053575. We also benefitted
from the High Performance Computing facilities at the School of Electrical Engineering
and Computer Sciences (SEECS) of the National University of Sciences and Technology (NUST),
Pakistan.  I.G. acknowledges support from the  Rustaveli National Science Foundation  No. 03/79.

\thispagestyle{empty}


\end{document}